\documentclass[%
 reprint,
superscriptaddress,
 amsmath,
 amssymb,
 aps,
]{revtex4-2}

\usepackage{graphicx}
\usepackage{dcolumn}
\usepackage{bm,amsmath}
\usepackage{hyperref}

\hypersetup{
    colorlinks,
    citecolor=blue,    filecolor=blue,
    linkcolor=blue,    urlcolor=blue
}

\begin{document}

\title{Ultra-Short Flying Focus}
\author{J\'er\^ome Touguet}
\affiliation{Laboratoire d’Optique Appliqu\'ee (LOA), CNRS, Ecole polytechnique, ENSTA, Institut Polytechnique de Paris, Palaiseau, France}
\author{Igor A. Andriyash}
\affiliation{Laboratoire d’Optique Appliqu\'ee (LOA), CNRS, Ecole polytechnique, ENSTA, Institut Polytechnique de Paris, Palaiseau, France}
\author{Lucas Rovige}
\affiliation{Laboratoire d’Optique Appliqu\'ee (LOA), CNRS, Ecole polytechnique, ENSTA, Institut Polytechnique de Paris, Palaiseau, France}
\author{C\'edric Thaury}
 \email{cedric.thaury@ensta.fr}
\affiliation{Laboratoire d’Optique Appliqu\'ee (LOA), CNRS, Ecole polytechnique, ENSTA, Institut Polytechnique de Paris, Palaiseau, France}

\begin{abstract}
Achromatic flying-focus enables programmable control of intensity peak velocity, with applications in ultrafast optics. However, spatiotemporal coupling inherently elongates ultrashort pulses by introducing frequency-dependent focusing and arrival-time dispersion. We present a theoretical model identifying this pulse-lengthening effect and propose a radially-dependent spectral chirp to compensate for chromatic timing mismatches. Numerical simulations confirm that this approach preserves both pulse duration and programmed flying-focus velocity over extended focal regions. Additionally, dispersive media such as plasmas can naturally mitigate elongation. These results extend achromatic flying-focus techniques to ultrashort pulses, enabling new opportunities in laser–plasma interactions and high-field nonlinear optics.
\end{abstract}

\maketitle

\section{Introduction}
The flying focus (FF) is a laser technique that uses spatiotemporal shaping to control the velocity of the laser intensity peak, enabling tunable subluminal, superluminal, or even backward motion ~\cite{Sainte-Marie:17, froula2018,clement,PhysRevLett.124.134802}. In a \textit{chromatic} flying focus~\cite{Sainte-Marie:17, froula2018}, a chirped laser pulse is focused by a chromatic optic, causing different frequency components to focus at different longitudinal positions and times. By adjusting the pulse chirp, the apparent focal position can be made to move at a widely tunable velocity along the optical axis. 

Conversely, an \textit{achromatic} flying focus can be achieved using a reflective focusing optic that generates a longitudinally extended focal line, such as an axiparabola~\cite{smartsev:19}, 
This optic establishes a one-to-one correspondence between the radial position $r$ on the mirror and the corresponding longitudinal focal point $z(r)$. By introducing a controlled radial delay $\tau(r)$ across the beam, the arrival time of the annular beamlets on axis, and hence the apparent velocity of the intensity peak, can be adjusted, while all frequencies, to first approximation, share the same extended focus set by the reflective geometry~\cite{2022JOpt...24d5503O,Liberman:24}. 

Both approaches enable programmable focal velocities decoupled from the group velocity and focal regions extending far beyond the Rayleigh length. However, while the chromatic flying focus requires a chirp that temporally stretches the pulse, limiting its suitability for ultrashort pulses, the achromatic flying focus can, in principle, operate with arbitrarily short pulses.

These capabilities have led to a growing number of applications across laser–plasma and ultrafast optics~\cite{Piccardo:25}. Chromatic flying-focus  have been used to better synchronize optical and X-ray pulses in soft X-ray lasers, resulting in shortest X-ray pulses~\cite{kabacinski2023}, and steer laser-produced THz emission in air~\cite{PhysRevLett.134.045001}. Among other applications, the achromatic FF  has been proposed for dephasingless laser–wakefield acceleration~\cite{PhysRevLett.124.134802,clement} control of two-color THz generation~\cite{PhysRevResearch.6.013041}, or photon acceleration~\cite{PhysRevA.104.043520}. 
For some applications, such as dephasingless plasma acceleration, the acceleration efficiency scales inverse to the pulse duration, which gives a strong motivation to use ultrashort laser pulses ~\cite{clement}. However, this benefit can only be realized if the pulse maintains its duration during propagation in the plasma.


In this work, we demonstrate that achromatic flying-focus pulses inherently undergo temporal elongation, even during propagation in vacuum, as the radial delay necessary to control propagation velocity translates into frequency-dependent focusing along the focal line. We show that this temporal elongation can be compensated by adding tailored radial chirp, and we numerically validate that our approach preserves the ultrashort pulse duration. These findings open the path to extending flying-focus techniques to the few-cycle regime, enabling new applications in compact particle acceleration and high-field nonlinear optics.

\section{Temporal Broadening in Free-Space Propagation}
To understand the origin of this temporal elongation, we first analyze pulse propagation in free space. The radial delay $\tau(r)$ required to synthesize an achromatic flying focus imposes a spatially dependent temporal shift on the pulse. The electric field of the collimated beam can therefore be written as
\begin{align}
    \mathbf{E}(r,t) = e^{-i\omega_{0} t}\,\mathbf{A}(t - \tau(r)),
\end{align}
where $r$ is the transverse coordinate, $\omega_{0}$ the carrier frequency, and $\mathbf{A}(t)$ the complex envelope.

Taking the Fourier transform yields the spectral field
\begin{align}
    \tilde{\mathbf{E}}(r,\omega)
    = e^{i\,\delta\omega\,\tau(r)}\,\tilde{\mathbf{A}}(\delta\omega),
\end{align}
with $\delta\omega = \omega - \omega_{0}$.  
This expression shows that the radial delay introduces a frequency-dependent phase $\phi_{stc}(r,\omega) = \delta\omega\,\tau(r)$, which couples frequency and transverse position.
The associated spectral phase can be interpreted as a wavefront curvature through
\begin{align}
    \phi_{stc}(r,\omega)
    =  \frac{\omega\, r^{2}}{2c\,R_c(r,\omega)},
\end{align}
where $R_c(r,\omega)= \omega\,r^2 /(2c\,\tau(r)\,\delta\omega)$
is the curvature radius associated with the frequency-dependent phase. The beam remains collimated at the central frequency ($R_c(r,\omega_0)=\infty$), while for $\omega \neq\omega_0$, the local curvature radius $R_c(r,\omega)$ is finite and inversely proportional to $\tau(r)\delta\omega$. Across the full spectral bandwidth $\Delta\omega$, the beam remains quasi-collimated if $\tau(r)\Delta(\omega)\ll  r\omega /c$.

After reflection by an axiparabola with local curvature $R_m(r)$, the curvature transforms, within the geometrical optics approximation, according to
\begin{align}
    R'(r,\omega)
    = \frac{R_c(r,\omega) \,R_m(r) }{2R_c(r,\omega)-R_m(r)}.
\end{align}
For a quasi-collimated beam ($ R_m(r) \ll R(r,\omega)$)
,  the focal position, measured as the distance to the mirror,  is located at
\begin{align}
    z(r,\omega) =  R'(r,\omega) \simeq f(r)
    + 2c\,\tau(r) \,\frac{\delta\omega}{\omega}\,\dfrac{f^{2}(r)}{r^2},
    \label{eq:z}
\end{align}
where $f(r)=R_m(r)/2$ denotes the focal distance of the axiparabola when no radial delay is applied.
Eq.~(\ref{eq:z}) shows that the effective focusing position of the axiparabola becomes not only radially dependent, through $R_m(r)$, but also frequency dependent due to the imposed radial delay. 
Consequently, when a radial delay ($\tau(r)\neq0$) is applied, rays originating from different radial positions $r$ can focus at the same longitudinal position $z$, but for different frequencies. Since these rays travel along different optical paths, they arrive at $z$ at slightly different times, resulting in temporal broadening of the pulse.
This effect becomes increasingly problematic as pulses grow shorter, both because their broader spectra enhance the variation in arrival times and because any temporal elongation represents a larger fraction of the pulse duration. 

To be more quantitative, we focus on a configuration in which an axiparabola generates a constant on-axis intensity from a flat-top beam of radius $R$, and first consider the case of quartic radial delay $\tau(r)=\beta r^4$. The axiparabola has an initial focal length $f_0$ and a focal depth $\delta_0$, with the focal position varying as $f(r) = f_0 + \delta_0\, r^2 / R^2$.  In the absence of radial delay, the on-axis intensity peak propagates superluminally in vacuum and, for $\delta_0\ll f_0$, its apparent velocity varies linearly with $z$ to first order, according to $v/c\approx 1+(z-f_0) (R^2/f_0\delta_0)$. Introducing a quartic radial delay modifies this slope and can, in particular, be used to achieve  propagation at a constant velocity~\cite{2022JOpt...24d5503O}.

Assuming $\delta_0 \ll f_0$ and propagating a distance $z$ in the Fresnel diffraction regime, the total phase accumulated by the pulse can be approximated as~\cite{smartsev:19,2022JOpt...24d5503O}:
\begin{align}
    \Phi(z,r,\omega)=\omega \dfrac{z}{c}+ \omega\beta_0 r^2\left(r^2-2r_0^2\right)+\phi_{stc}(r,\omega) +\mathcal{O}(r^6),
    \label{eq:phase}
\end{align} 
where $\beta_0=\delta_{0} / 4cf_{0}^{2}R^{2}$ characterizes the quadratic radial phase curvature induced by the axiparabola, $r_0^2=(z-f_0) / 4\beta_0cf_0^2$, and $r$ is the radial coordinate at $z=0$.
At a given propagation distance $z$, the dominant radial contribution originates from an annular beamlet with radius $r_s$, determined by the stationary-phase condition
$\partial\Phi / \partial r=0$, yielding
\begin{align}
    r_s^2= r_0^2\left(1+\dfrac{\beta}{\beta_0}\dfrac{\delta\omega}{\omega}\right)^{-1}.
\end{align}
which directly identifies $r_0$ as the stationary radius in absence of radial delay. 
The group delay~\cite{DIELS200661} can then be computed by substituting $r_s$ into Eq.~(\ref{eq:phase}):
\begin{align}
    t_g(z,\omega) &= \dfrac{\partial\phi}{\partial\omega}(z,r_s,\omega) \\
    &= \dfrac{z}{c}+(\beta-\beta_0)r_0(z)^4 \nonumber\\
    &\phantom{=} - 2\dfrac{\beta^2}{\beta_0}r_0(z)^4\dfrac{\delta\omega}{\omega_0}
    + \mathcal{O}\left(\dfrac{\delta\omega^2}{\omega_0^2}\right)
    \label{eq:group_delay}
\end{align}
Through the $z$-dependence of $r_0$, the second term in Eq.~(\ref{eq:group_delay}) alters the on-axis arrival time and thus the velocity of the intensity peak. At the central frequency, this effect vanishes for $\beta=\beta_0$, yielding propagation at the speed of light $c$. The third term introduces spatially dependent group-delay dispersion (GDD), causing a temporal chirp and pulse broadening that increase with propagation distance. For ultrashort laser pulses, higher-order terms in the Taylor expansion of Eq.~(\ref{eq:group_delay}) may become significant, resulting in a nonlinear chirp.

\section{Pulse-Broadening Mitigation}
\begin{figure}
    \centering
    \includegraphics[width=1\linewidth]{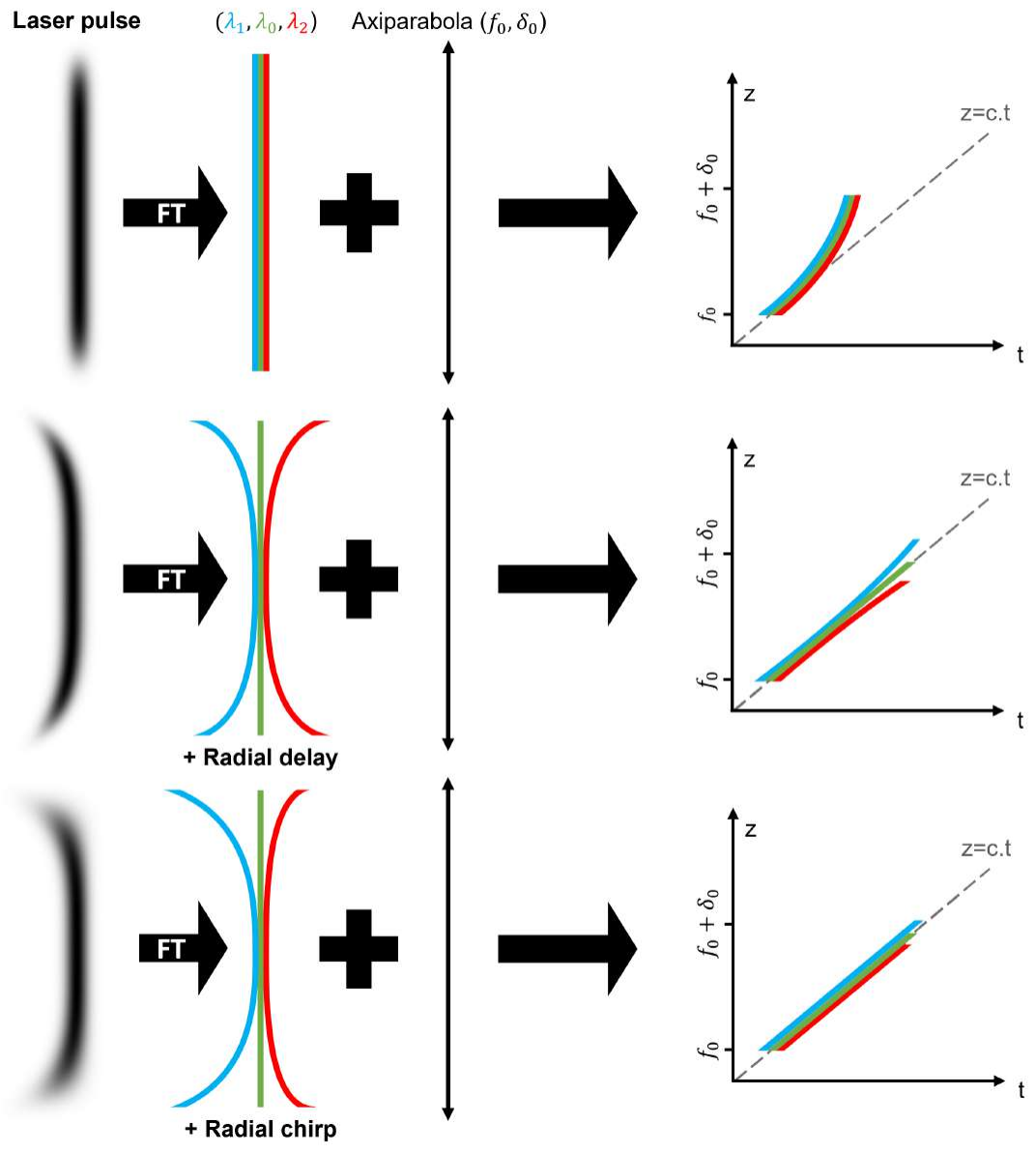}
    \caption{Mitigation of pulse broadening at the focus of an axiparabola for different wavelengths.
   Without any radial delay or chirp (top), the pulse propagates along the z-axis at a superluminal speed, deviating from the  $z=ct$ trajectory. Adding a quartic radial delay (middle) compensates this temporal shift at the central frequency, chromatic effects lead to pulse stretching along the focal line. Adding both a radial delay and a radial chirp (bottom) aligns all spectral components  along the $z=ct$ trajectory, preserving the pulse duration while maintaining the programmed flying-focus velocity.}
    \label{fig:fig1}
\end{figure}

To mitigate this GDD-induced pulse broadening, we can introduce a radial chirp delay to the radial delay $\tau(r)$:
\begin{align}
    \tau(r,\omega)=\beta r^4+\dfrac{\beta^2}{\beta_0}\dfrac{\delta\omega}{\omega_{0}}r^4.
\end{align}
By canceling the chromatic timing mismatch between spectral components, this radial chirp preserves the pulse duration while retaining the required flying-focus velocity, as illustrated in Fig~\ref{fig:fig1}.

\begin{figure}
    \centering
\includegraphics[width=0.8\linewidth]{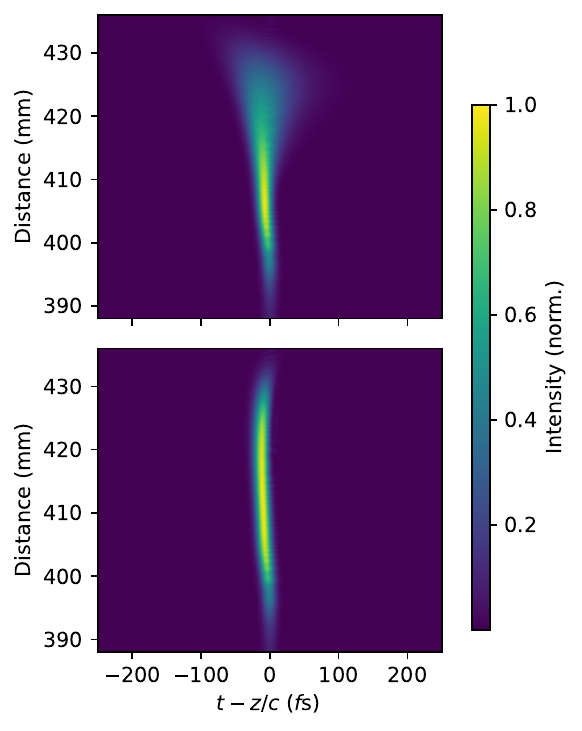}
    \caption{Simulated on-axis intensity in the frame of a particle moving at speed $c$. 
    Without radial chirp (top), the imposed radial delay yields a luminal apparent propagation velocity but leads to progressive pulse stretching and reduced peak intensity. With radial chirp (bottom), the pulse duration is preserved along the entire focal line while maintaining a luminal apparent velocity.
    }
    \label{fig:fig2}
\end{figure}

We used the \textit{Axiprop} optical propagation Python package \cite{axiprop_article} to produce a focal line of constant intensity and length $\delta_0 = 30$~mm, starting at $z = f_0 = 400$~mm and $R=50$~mm. They are performed both with and without radial chirp, for a pulse duration of $\tau_{las}=$10 fs, defined as the 1/e temporal half-width of the electric-field envelope, and  $\beta_0=\,0.57$~fs.cm$^{-4}$. 
Figure~\ref{fig:fig2} shows the evolution of the on-axis intensity in the reference frame of a particle moving at speed $c$. In this frame, luminal propagation appears as a vertical trajectory, while deviations from $c$ manifest as temporal displacements along the $z-ct$ axis. In the absence of radial chirp, the imposed radial delay produces a luminal apparent velocity, but the pulse gradually elongates along the focal line, reaching a pulse duration $\tau_{s}> $ 40 fs and reducing its peak intensity. When the radial chirp is introduced, the pulse duration is preserved and the apparent velocity remains luminal. Importantly, the stretching does not depend on the method used to introduce the spatio-temporal couplings required for the flying focus.

While the quartic radial delay effectively modifies the slope of the velocity profile, many applications require additional degrees of freedom, particularly the ability to tune the initial velocity. This can be achieved by superimposing a quadratic delay. We therefore generalize the radial delay as follows:
\begin{align}
    \tau(r) = \alpha\,r^{2} + \beta\,r^{4},
\end{align}
with the detailed calculation of the resulting group delay provided in Appendix~\ref{app:A}. This additional term modifies the focal shift, introducing a frequency-dependent temporal shift along the propagation axis:
\begin{align}
    \Delta\tau(z,\delta\omega)
    \approx&
   -\left( \dfrac{\alpha^{2}}{4}    +\alpha\beta\,r_0^{2}\    +\beta^2\,r_0^{4}\right) \,\dfrac{2}{\beta_{0}} \,\dfrac{\delta\omega}{\omega_{0}}\nonumber \\
   &+\mathcal{O}\left(\dfrac{\delta\omega^2}{\omega_{0}^2}\right).
    \label{eq:2nd/4th order}
\end{align}
We identify three distinct contributions to the temporal stretching: a spatially uniform term corresponding to a global temporal chirp, a quadratic radial chirp, and a quartic radial chirp. The quadratic term results from the coupling between the quadratic and quartic components of the imposed radial delay. We then derive the radial delay required to compensate for the induced pulse stretching:
\begin{align}
\tau(r,\omega)=
\overbrace{\alpha\,r^2+\beta\,r^4}^{\text{FF radial delay}}
+\overbrace{\left(\dfrac{\alpha^2}{4}+\beta\alpha r^2+\beta^2 r^4\right)\dfrac{1}{\beta_0}\dfrac{\delta\omega}{\omega_0}}^{\text{Elongation compensation}}
\label{eq:delay_full_vac}
\end{align}

\begin{figure}
    \centering
    
\includegraphics[width=0.8\linewidth]{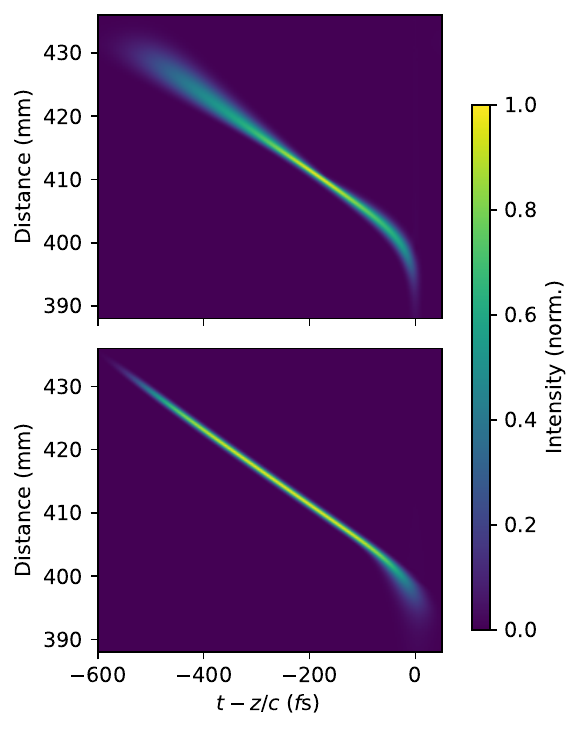}
    \caption{
    Simulated on-axis intensity in the frame of a particle moving at speed $c$. 
    Quartic chirp alone (top) leads to alternating temporal stretching and partial compression along the focal line, with a compression point near mid-propagation. Quartic chirp combined with additional global and quadratic chirp (bottom) maintains temporal compression along the full focal line.}
    \label{fig:fig3}
\end{figure}

We verified the effectiveness of this radial delay using numerical simulations with the same parameters as in Fig.~\ref{fig:fig2}, but including quartic chirp, $\alpha = -20$~fs.cm$^{-2}$ and the radial chirp given by Eq.~(\ref{eq:delay_full_vac}). Figure~\ref{fig:fig3} compares the resulting on-axis intensity evolution with and without a global chirp.

In the absence of additional chirp, the pulse initially undergoes temporal stretching along the focal line. After propagating approximately 10~mm, it becomes compressed before stretching again toward the end of the line. This behavior arises because the global and quadratic chirps have opposite signs, resulting in partial compensation near the midpoint of the propagation distance.
By contrast, when an additional chirp is introduced, the pulse remains temporally compressed throughout the entire propagation distance, as the added chirp effectively compensates for the combined effects of the global and quadratic contributions.
A slight elongation, attributed to the global chirp, is observed only in the intensity up-ramp ($z<f_0$), a region where the intensity is inherently low.

\section{Generation of Radially-Dependent Spectral Chirp}
A radially dependent delay is a key requirement for flying-focus beams, and several approaches can 
be considered to generate it. 
Reflective solutions, such as radial echelons~\cite{Pigeon:24}, provide a straightforward approach compatible with high-intensity laser systems. 
Chromatic refractive~\cite{Jolly_2020,Smartsev_2022,Sainte-Marie:17} or diffractive~\cite{2018NaPho..12..262F} optics can also produce such delays via longitudinal chromatism in the far field, which translates into a radial group delay in the near field. When placed in non-collimated regions, these elements provide tunability~\cite{Kabacinski_2021}. 
More flexible schemes based on spatial light modulators and deformable mirrors~\cite{2020CmPhy...3..211L,Ambat2023} allow programmable control of phase and radial delay, at the expense of a significantly reduced damage threshold.

By contrast, the generation of a radial chirp remains largely unexplored, and practical solutions have yet to be identified. 
Radial chirp naturally arises in transmissive optics through higher-order dispersion of the refractive index, but exploiting this effect in a controlled manner is challenging. For reflective optics, it has been suggested that radial chirp can be introduced by applying a variable-thickness coating to the surface of the echelon \cite{Miller2023}. 
In principle, a radial chirp could also be generated by mapping wavelength onto the radial coordinate using a compressor-based setup and applying wavelength-dependent phases accordingly~\cite{2023NaPho..17..822P}; however, implementing such a mapping in a stable and controllable manner is highly nontrivial.

Here, we propose a practical solution 
based on dispersive transmissive optics acting as a radial echelon that can induce both radial delay and radial chirp necessary for ultra-fast flying-focus. 
One way to realize this concept is to combine two optical materials with different dispersive properties. Let the materials have refractive indices $n_{1}(\omega)$ and $n_{2}(\omega)$, where one is weakly dispersive and the other strongly dispersive. By fabricating two radially varying thickness profiles, $L_{1}(r)$ and $L_{2}(r)$, the doublet ensures that different frequencies accumulate different optical path lengths as they propagate and for a given frequency $\omega$, the total optical path is
\begin{align}
    \mathrm{OPL}(\omega,r)
    = \left(n_{1}(\omega)-1\right)\,L_{1}(r)
    + \left(n_{2}(\omega)-1\right)\,L_{2}(r)
    + L_{\text{tot}},
\label{eq:OPL}
\end{align}
where $L_{\text{tot}}$ is a reference path length. 
The objective is to design the thickness profiles $L_{1}(r)$ and $L_{2}(r)$ such that the optical path length for the central frequency $\omega_{0}$ yields a general radial delay $\tau_{d,0}(r)$, while a neighboring frequency $\omega_{0}+\Delta\omega$ experiences an additional prescribed radial delay $\tau_{d,0}(r)+\tau_{d,1}(r)$. Specifically, we impose
\begin{align}
\begin{cases}
\mathrm{OPL}(\omega_{0},r)
=
\mathrm{OPL}(\omega_{0},0)+\,c \tau_{d,0}(r)
\\[6pt]
\mathrm{OPL}(\omega_{0}+\Delta\omega,r)
=
\mathrm{OPL}(\omega_{0}+\Delta\omega,0)
+
c\left(\tau_{d,0}(r)+\tau_{d,1}(r)\right).
\end{cases}
\label{eq:eqsys}
\end{align}
Combining Eq.~(\ref{eq:OPL}) and (\ref{eq:eqsys}) provide two linear constraints on the two thickness profiles $L_1(r)$ and $L_2(r)$.
\begin{figure}
    \centering

\includegraphics[width=0.7\linewidth]{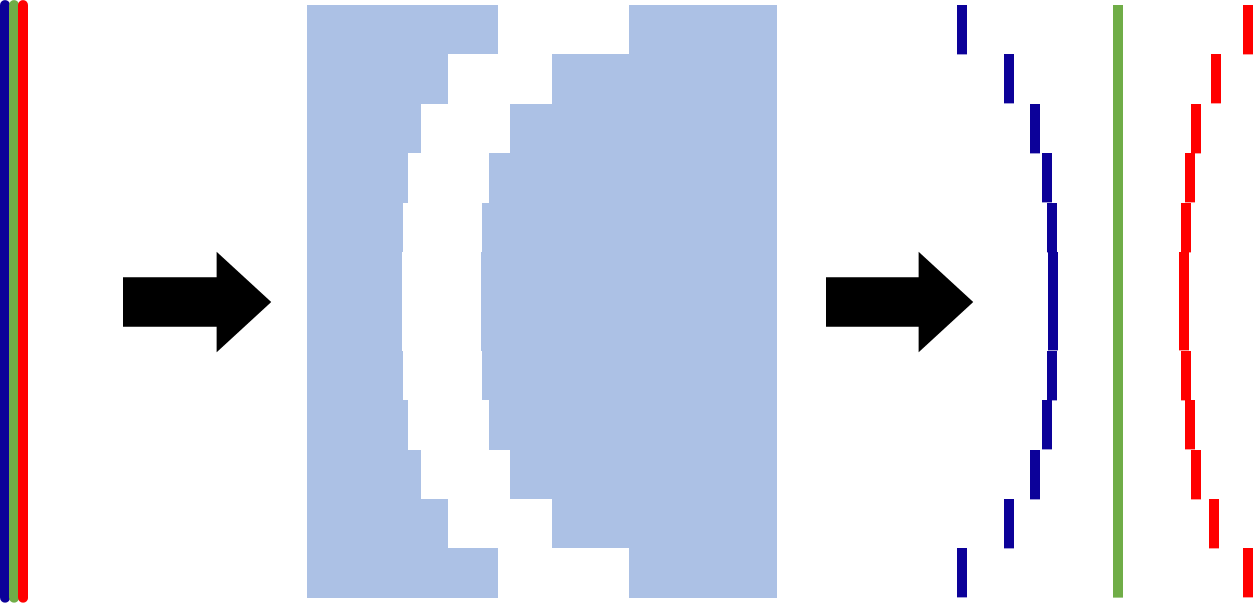}
    \caption{Illustration for different wavelengths of a stepped transmissive doublet for generating radially-dependent spectral chirp.}
    \label{fig:fig4}
\end{figure}
In practice, implementing such profiles with continuous surfaces would introduce unwanted focusing and spherical aberrations due to residual curvature. To avoid these effects, we adopt a stepped transmissive doublet, in which the thickness varies discretely in concentric annular zones, as illustrated in Fig.~\ref{fig:fig4}.

For annular steps of height $h$, and assuming quadratic dispersion of the optical index~\cite{cauchy_law}, the phase imparted by the stepped doublet is
\begin{align}
    \phi^d_{stc}(r,\omega)
    = \dfrac{\omega}{c} \,h\left\lfloor \dfrac{c\,\tau_{d,0}(r)}{h} \right\rfloor\left(1+\gamma\frac{\delta\omega}{\omega_0}\left(1+\frac{\delta\omega}{2\omega_0}\right)\right),
\end{align}
with $\lfloor \,\,\rfloor$ the floor function and $\gamma=\tau_{d,1}(r)/\tau_{d,0}(r)(\Delta\omega/\omega_0+\Delta\omega^2/2\omega_0^2)$. Choosing a step height $h = \lambda_0$ ensures that, at $\omega_0$, each step introduces an integer multiple of $2\pi$ phase, rendering the phase uniform modulo $2\pi$. The leading-order behavior of the imparted phase is therefore governed by its frequency dependence, and expanding in $\delta\omega/\omega_0$ gives
\begin{align}
    \phi^d_{stc}(r,\omega)
    \approx &
    \phi^d_{stc}(0,\omega_0)\nonumber\\
    &+\delta\omega\,\tau_{d,0}(r)
    \left(
    1 +\gamma+ \gamma\frac{3\delta\omega}{2\omega_0}
    + \mathcal{O}\!\left(\frac{\delta\omega^2}{\omega_0^2}\right)
    \right).
\end{align}

\begin{figure}
    \centering

\includegraphics[width=0.8\linewidth]{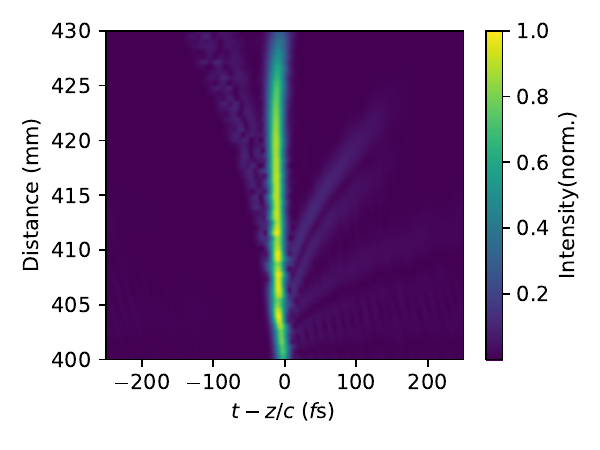}
    \caption{Simulated on-axis intensity in the frame of a particle moving at speed $c$, using a stepped doublet with radial delay $\tau_d(r)=\beta_0r^4/3$ and $\gamma=2$. The pulse duration is preserved along the entire focal line while maintaining a luminal apparent velocity.}
    \label{fig:fig4b}
\end{figure}

This phase corresponds to a prescribed radial group delay $\tau(r,\omega)=\tau_{d,0}(r)(1+\gamma+3\gamma\delta\omega/2\omega_0)$. In particular, choosing $\tau_{d,0}(r)=\beta_0r^4/3$ and $\gamma=2$, yields luminal apparent velocity while preserving the pulse duration along the focal line. Numerical simulations using the discretized phase $\Phi_d$ show that the pulse continues to propagate at $c$ with minimal temporal broadening in Figure~\ref{fig:fig4b}. Diffraction artifacts arising from the radial discretization are visible, but they remain weak and do not significantly perturb the intended flying-focus dynamics.

\section{Propagation in a Dispersive Medium}

Beyond the pulse broadening inherent to vacuum propagation, additional chromatic effects can arise when the beam propagates through a dispersive medium, such as a plasma. When combined with an axiparabola, these effects introduce a radially dependent dispersion. 
To illustrate this phenomenon, we consider a beam focused by an axiparabola as it propagates through a plasma region, modeled here as a cube with an optical index given by $\eta(\omega)=\sqrt{1-\dfrac{\omega_p}{\omega}}$, beginning at $f_0$.
Using the same approach as for vacuum propagation, we compute the group delay in the plasma, including refraction and dispersion. Neglecting the radial delay yields (see Appendix B):
\begin{align}
    t_g(z,\omega)\approx&\dfrac{z}{c}-\alpha_0r_0(z)^2+2\alpha_0r_0(z)^2\dfrac{\delta\omega}{\omega_0}\nonumber\\
    &-\beta_0r_0(z)^4+\mathcal{O}\left(\dfrac{\delta\omega^2}{\omega_{0}^2}\right),
    \label{eq:chirp_plasma}
\end{align}
with $\alpha_0=(\eta-1)\delta_0/cR^2$. We observe an elongation 
similar to that induced by a quadratic radial chirp. Moreover, considering a radial delay $ \tau(r) = \alpha\,r^{2} + \beta\,r^{4}$ and combining Eqs.~(\ref{eq:chirp_plasma}) and~(\ref{eq:2nd/4th order}), the group delay can be expressed as:
\begin{align}
    t_g(z,\omega)\approx&\dfrac{z}{c}+(\alpha-\alpha_0)r_0^2+(\beta-\beta_0)r_0(z)^4 \, \nonumber \\
    & -\left( \dfrac{\alpha^{2}}{4}    +(\alpha\beta-\alpha_0\beta_0)\,r_0^{2}\    +\beta^2\,r_0(z)^{4}\right) \,\dfrac{2}{\beta_{0}} \,\dfrac{\delta\omega}{\omega_{0}}\nonumber \\
   & +\mathcal{O}\left(\dfrac{\delta\omega^2}{\omega_{0}^2}\right).
    \label{eq:delay_plasma}
\end{align}
The case $\alpha = \alpha_0$ and $\beta = \beta_0$ is of particular interest
, as it leads to propagation at $c$ in the plasma, which is a necessary condition for dephasing-free laser–plasma acceleration~\cite{clement}. Under these conditions,
Eq.~(\ref{eq:delay_plasma}) simplifies to

\begin{equation}
t_g(z,\omega) \approx \frac{z}{c} - \frac{2}{\beta_0} \left( \frac{\alpha_0^2}{4} + \beta_0^2 r_0(z)^4 \right) \frac{\delta \omega}{\omega_0} + \mathcal{O}\Big(\frac{\delta \omega^2}{\omega_0^2}\Big).
\end{equation}
During propagation, the plasma-induced pulse elongation is exactly compensated by the quadratic radial chirp arising from the combined quadratic and quartic radial delay. This balance allows the pulse to remain temporally compressed using only global and quartic chirp.

\begin{figure}
    \centering

\includegraphics[width=0.8\linewidth]{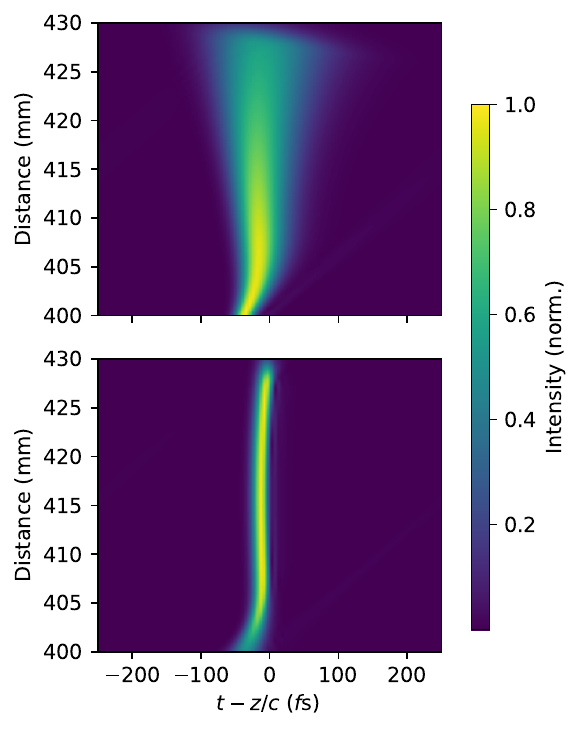}
    \caption{Simulated on-axis intensity in a plasma in the frame of a particle moving at speed $c$. In the absence of chirp (top), the imposed radial delay yields a luminal apparent propagation but elongated and leads to progressive pulse stretching and reduced peak intensity. Adding global and quartic chirp (bottom) compensates this temporal elongation, maintaining pulse compression along the entire focal line while keeping luminal propagation}
    \label{fig:fig5}
\end{figure}

Figure~\ref{fig:fig5} illustrates this behavior through simulations performed in a plasma with an electron density \(n_e = 1.7 \times 10^{19}~\mathrm{cm^{-3}}\), using the same axiparabola and laser pulse parameters as in vacuum. 
In the absence of radial chirp (top panel), the pulse stretches rapidly at first and then continues to elongate more gradually during propagation.
This elongation 
can be quantified for a Gaussian transform-limited pulse (\(\sigma_\omega = 2/\tau_{\rm las}\)) as
\begin{align}
\tau_{s}(z) = \left[\tau_{\rm las}^2 + \left( \frac{\alpha_0^2}{\beta_0 \omega_0 \tau_{\rm las}} + \frac{4 \beta_0}{\omega_0 \tau_{\rm las}} r_0(z)^4 \right)^2 \right]^{1/2}.
\label{eq:elongation}
\end{align}

According to Eq.~(\ref{eq:elongation}), the pulse duration reaches \(\tau_{s} > 80~\mathrm{fs}\) by the end of propagation, eight times its initial value, in agreement with the simulation results. This strong temporal broadening leads to a substantial reduction of the peak intensity, thereby significantly degrading the interaction efficiency. However, when a global and quartic radial chirp is applied, the plasma allows the laser pulse to propagate along the axis at the speed of light while remaining temporally compressed, as shown in the bottom panel of Fig.~\ref{fig:fig5}.

\begin{figure}
    \centering
\includegraphics[width=0.9\linewidth]{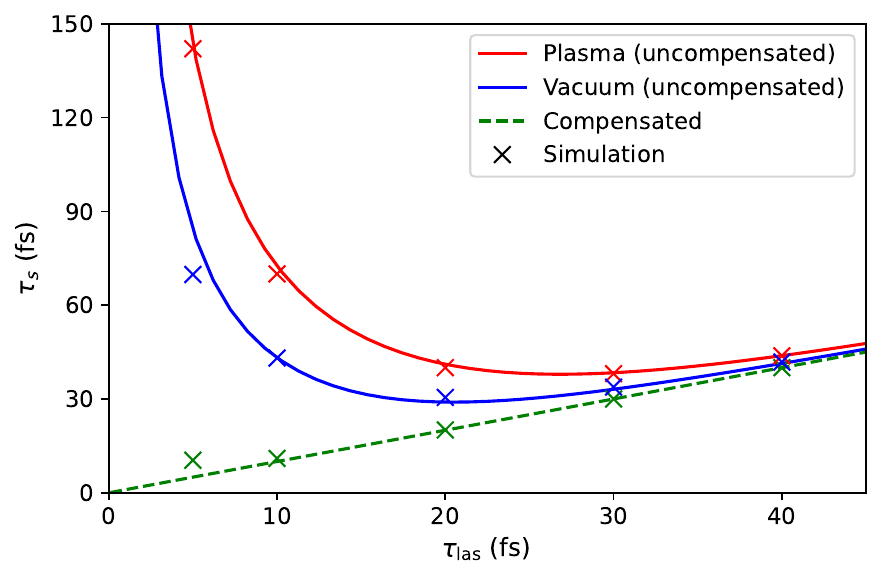}
    \caption{Stretched pulse duration $\tau_s$ as a function of the initial laser duration $\tau_{las}$ in plasma ($n_e = 1.7 \times10^{19}$ cm$^{-3}$) and vacuum with a luminal group velocity. Cases without radial chirp are shown in red (plasma) and blue (vacuum), while the compensated case with radial chirp is shown in green. Symbols correspond to simulation results and lines indicate the analytical prediction.}
    \label{fig:fig7}
\end{figure}

Figure~\ref{fig:fig7} shows the dependence of the stretched pulse duration $\tau_s$ on the initial laser duration $\tau_{las}$ in plasma assuming an imposed luminal group velocity. For comparison, we also include the case of an imposed luminal group velocity in vacuum ($\tau(r)=\beta_0r^4$). The stretched duration $\tau_s$ is evaluated near the end of the focal line at $z = 425$~mm. Shorter pulses undergo stronger elongation, and the results closely follow the analytical prediction given by Eq.~(\ref{eq:elongation}).  For $\tau_{las} = 5~\mathrm{fs}$, noticeable deviations from the analytical model emerge. In this regime, higher-order terms in the group delay must be included in the series expansion to accurately reproduce the numerical results.

\section{Conclusion}

We have analyzed the spatiotemporal dynamics of achromatic flying-focus pulses, identifying an intrinsic distortion that causes temporal elongation, even in vacuum, due to frequency-dependent focusing. This effect is particularly critical for ultrashort pulses. By introducing a frequency-dependent radial delay, we fully compensate for this distortion, enabling propagation of temporally compressed pulses over extended focal regions while preserving both pulse duration and programmed focal velocity, as confirmed by numerical simulations.

These results provide a practical route to ultrashort achromatic flying-focus pulses, overcoming a key limitation for applications requiring simultaneous control of focal velocity and temporal structure. The method also extends to plasma-based scenarios, where spatio-temporal coupling control is crucial for optimizing laser-plasma interactions, such as in wakefield acceleration. This advancement unlocks new possibilities for high-precision, high-energy applications in ultrafast optics and advanced acceleration.

\bibliography{biblio}

\newpage
\appendix
\section{Calculation of the group delay for quadratic and quartic radial delays}
\label{app:A}
In the limit $\delta_0\ll f_0$, the phase applied by an axiparabola is $$ \Phi(r,\omega)\approx kr^2\left(\dfrac{1}{2f_0}-\beta_0c\,r^2\right)\mathrm{,}$$
with $\beta_0=\delta_{0}/4cf_{0}^{2}R^{2}$. 
The phase applied in the Fresnel diffraction regime to the pulse at the position $z$ is then given by:
\begin{align}
    \Phi(r,\omega)\approx\omega \dfrac{z}{c}+ \omega\beta_0 r^2\left(r^2-2r_0^2\right)+\delta\omega\tau(r) +\mathcal{O}(r^6),
    \label{eqA:phase}
    \end{align}
with $r_0^2=\dfrac{(z-f_0)}{4\beta_0\,f_0^2\,c}$. For a given position z, the dominant radial contribution is determined by the stationary phase condition  $\partial\Phi/\partial r(z,\omega)=0$. This yields a stationary radius:
\begin{align}
    r_s^2&=\dfrac{r_0^2\left(1-2c\,\alpha\dfrac{\delta\omega}{\omega}\dfrac{f_0^2}{z-f_0}\right)}{1+\dfrac{\beta}{\beta_0}\dfrac{\delta\omega}{\omega}}.
\end{align}
We can substitute this stationary radius into Eq.~(\ref{eqA:phase}) to obtain an expression for the phase at position $z$:
\begin{align}
    \Phi(z,\omega)&\approx\omega\dfrac{z}{c}-\omega\beta_0r_0^4+\delta\omega(\alpha r_0^2+\beta r_0^4)
    \nonumber\\
    &-\left(\dfrac{\alpha^2}{4c\beta_0}+\dfrac{\beta\alpha}{\beta_0}r_0^2+\dfrac{\beta^2}{\beta_0}r_0^4\right)\dfrac{\delta\omega^2}{\omega_0}+\mathcal{O}\left(\dfrac{\delta\omega^3}{\omega_{0}^2}\right)
\end{align}
From the phase, the group delay can be determined through:
\begin{align}
    t_g(z,\omega)&=\dfrac{\partial\Phi}{\partial\omega}(z,\omega\nonumber)\\
    &\approx\dfrac{z}{c}+(\beta-\beta_0)r_0^4+\alpha r_0^2
    \nonumber\\
    &-\left(\dfrac{\alpha^2}{2c\beta_0}+2\dfrac{\beta\alpha}{\beta_0}r_0^2+2\dfrac{\beta^2}{\beta_0}r_0^4\right)\dfrac{\delta\omega}{\omega_0}+\mathcal{O}\left(\dfrac{\delta\omega^2}{\omega_{0}^2}\right).
\end{align}
 A radial chirp can be introduced in the delay profile to counteract this pulse stretching.
\begin{align}
    \tau(r,\omega)=\alpha\,r^2+\beta\,r^4+\left(\dfrac{\alpha^2}{4c\beta_0}+\dfrac{\beta\alpha}{\beta_0}r^2+\dfrac{\beta^2}{\beta_0}r^4\right)\dfrac{\delta\omega}{\omega_0}.
\end{align}


\section{Calculation of the group delay in a plasma}
\label{app:B}
Considering that we propagate in a medium of optical index $\eta=\sqrt{1-\dfrac{\omega_p^2}{\omega^2}}$ at position $z=f_0$ with $k_1=\eta\,k_0=\eta\dfrac{\omega}{c}$ and $\dfrac{\partial \omega
}{\partial k}=\eta\,c$.
The phase applied in the Fresnel diffraction regime to the pulse at the position $z=f_0$ is
\begin{align}
    \Phi(r,\omega)= k_0f_0+k_0\beta_0\, cr^4.
\end{align}
At the transition, the radius is reduced to $r'=(z-f_0)r/z$, and the phase at position $z>f_0$ is
\begin{align}
    \Phi(r,\omega)&=k_0f_0+k_1(z-f_0)- k_1\dfrac{r'^2}{2(z-f_0)}+k_0\beta_0\, cr^4\\
    &\approx k_0f_0+k_1(z-f_0)- k_1\dfrac{(z-f_0)}{2f_0^2}r^2+k_0\beta_0\, cr^4.
    \label{eqB:phase}
\end{align}
The dominant radial contribution is determined by the stationary phase condition  $\partial\Phi/\partial r(z,\omega)=0$  yielding:
\begin{align}
    r_s^2&\approx\eta\dfrac{z-f_0}{4\beta_0cf_0^2}.
\end{align}
We can substitute the stationary radius into Eq.~(\ref{eqB:phase}) to obtain an expression of the group delay:
\begin{align}
    t_g(z,\omega)&=\dfrac{\partial\phi}{\partial\omega}(z,\omega)\\
    &=\dfrac{z}{c}+\left(\dfrac{1}{\eta}-1\right)\dfrac{z-f_0}{c}+(\eta^2-2)\dfrac{(z-f_0)^2}{16\beta_0c^2 f_0^4}.
    \label{eqB:time_plasma_appendix}
\end{align}
Considering $\eta \ll1$, Eq.~(\ref{eqB:time_plasma_appendix}) simplifies to:
\begin{align}
    t_g(z,\omega)&\approx\dfrac{z}{c}-\left(\eta-1\right)\dfrac{z-f_0}{c}-\dfrac{(z-f_0)^2}{16\beta_0c^2 f_0^4}\\
    &\approx\dfrac{z}{c}-(\eta-1)\dfrac{\delta_0}{cR^2}r_0^2-\beta_0r^4.
    \label{eqB:time_plasma_appendix_simp}
\end{align}
Moreover, the optical index is dependent of the frequency and at first order, it derives to:
\begin{align}
    \eta(\omega)&\approx\eta_0+\delta\omega\left.\frac{d\eta}{d\omega}\right|_{\omega=\omega_0}\\
    &\approx\eta_0-2(\eta_0-1)\dfrac{\delta\omega}{\omega_0}.
    \label{eqB:eta_derivation}
\end{align}
So, combining Eq.~(\ref{eqB:time_plasma_appendix_simp}) and (\ref{eqB:eta_derivation}) gives a stretching of the pulse during the propagation:
\begin{align}
    t_g(z,\omega)&\approx\dfrac{z}{c}-\left(\eta_0-1\right)\dfrac{z-f_0}{c}-\dfrac{(z-f_0)^2}{16\beta_0c^2 f_0^4}\\
    &+2(\eta_0-1)\dfrac{z-f_0}{c}\dfrac{\delta\omega}{\omega_0}\\
    &\approx\dfrac{z}{c}-\alpha_0r_0^2-\beta_0r_0^4+2\alpha_0r_0^2\dfrac{\delta\omega}{\omega_0},
\end{align}
with $\alpha_0=(\eta_0-1)\dfrac{\delta_0}{cR^2}$.
\end{document}